\documentclass[pre,reprint,amsmath,amssymb]{revtex4-1}
\usepackage[english]{babel}
\usepackage[dvips]{graphicx}
\usepackage{units}

\newcommand{\e}{\operatorname{e}}

\renewcommand{\i}{i}

\newcommand{\E}[1]{$\times10^{#1}$}
\newcommand{\ten}[1]{$10^{#1}$}
\newcommand{\wcm}{~W\,cm$^{-2}$}
\newcommand{\ccm}{~cm$^{-3}$}

\newcommand{\mum}{~$\mu$m}

\newcommand{\etal}{\emph{et al.}}

\renewcommand{\c}[1]{}

\newcommand{\ifrac}[2]{(#1/#2)}

\newcommand{\dz}{\partial_z}
\newcommand{\dt}{\partial_t}

\newcommand{\real}{\mathfrak{Re}}

\begin{document}

\title{Raman amplification in the coherent wavebreaking regime}
\author{J.~P. Farmer}
\author{A. Pukhov}
\affiliation{Heinrich Heine Universit\"at, 40225 D\"usseldorf, Germany}

\begin{abstract}\noindent
  In regimes far beyond the wavebreaking threshold of Raman amplification, we show that significant amplifcation can occur after the onset of wavebreaking, before phase mixing destroys the coherent coupling between pump, probe and plasma wave.  Amplification in this regime is therefore a transient effect, with the higher-efficiency ``coherent wavebreaking'' (CWB) regime accessed by using a short, intense probe.  Parameter scans illustrate the marked difference in behaviour between below wavebreaking, in which the energy-transfer efficiency is high but total energy transfer is low, wavebreaking, in which efficiency is low, and CWB, in which moderate efficiencies allow the highest total energy transfer.
\end{abstract}

\maketitle

\section{Introduction}
Raman amplification in plasma has been suggested as a mechanism to allow the creation of ultrashort, ultraintense laser pulses, \cite{raman-shvets-compton} which have applications across science and technology.  Using plasma as a gain medium offers the advantage that, unlike solid-state media, it does not have a damage threshold, potentially reducing or removing the need for stretching and compression of the laser pulse, allowing either a reduction in size or an increase in power of next-generation laser systems.

In the Raman interaction, a probe pulse interacts with a counterpropagating pump, driving a plasma wave through the ponderomotive force of their beat.  If the pump and probe frequencies, $\omega_a$ and $\omega_b$, are chosen such that the probe is downshifted from the pump by the plasma frequency, $\omega_p$, the plasma wave is resonantly excited and may grow to large amplitude.  The resulting density perturbation acts as a moving Bragg grating, which acts to scatter and Doppler-shift the pump pulse into the probe, amplifying the latter.

Although the plasma itself does not have a damage threshold, there remain several effects which act to limit the interaction, such as filamentation and parasitic spontaneous backscatter of the pump \cite{raman-trines-simulation}.  Further, the high wavenumber and low phase velocity of the laser beat, which allows a large-amplitude plasma wave to be excited, means that the plasma wave can break for even moderate laser intensities, $\ll$ 1\E{18}\wcm\ \cite{raman-malkin-pumpdepletion}, leading to strong damping \cite{plasma-dawson-nonlinear}, which will limit the coupling between pump and probe \cite{raman-yampolsky-break}.

\c{While these effects have been the subject of numerous theoretical and computational works \cite{raman-trines-simulation,raman-toroker-break}, experimental campaigns continue to show significantly lower energy transfer efficiency than predicted \cite{raman-cheng-experiment,raman-vieux-experiment,raman-pai-experiment}, the best to date being by Ren~\etal, achieving 6.4\% in a double-pass setup \cite{raman-ren-experiment}.  There are many potential reasons for this discrepancy, from the idealised laser pulses interacting in idealised plasma used in simulations and theory, to the complexities of such an experiment.  However, it is interesting to note that efficiencies reported in simulations can vary significantly, notably the results of Trines~\etal\ \cite{raman-trines-simulation} and Toroker~\etal\ \cite{raman-toroker-break}.  This highlights the sensitivity of the interaction to seemingly minor details, which must be better understood if experimental works are to achieve sufficient amplification to be 
considered viable for applications.}

These effects likely contribute to the low efficiencies reported by experimental campaigns \cite{raman-balakin-break,raman-cheng-experiment,raman-pai-experiment,raman-vieux-experiment}, the best to date being by Ren~\etal, achieving 6.4\% in a double-pass setup \cite{raman-ren-experiment}.  Previous theoretical and computational works have focussed on the optimal pump amplitude for efficient amplification \cite{raman-yampolsky-break,raman-trines-simulation,raman-toroker-break,raman-edwards-efficiency}.  Although it is known that the highest efficiencies are achieved below the wavebreaking threshold, the necessarily low intensities would require large interaction volumes to scale to high power.  In this work we therefore focus on the energy-transfer efficiency far beyond the wavebreaking threshold.  

It is interesting to note, however, that simulations carried out in this regime have yielded significantly different efficiencies, even for similar parameters, ranging from 35\% in the work of Trines \etal \cite{raman-trines-simulation}, to less than 10\% in that of Toroker \etal \cite{raman-toroker-break}.  These apparently contradictory results have been the subject of significant discussion \cite{raman-toroker-break,raman-edwards-efficiency}, although the main cause has not previously been identified.  Clearly, reconciling the differences between simulations is important if reliable comparison to experiment is to be made.

In this work we identify a new process in the Raman interaction, coherent wavebreaking.  In section~\ref{probe}, we illustrate a strong dependence of the energy-transfer efficiency on the probe duration, which explains the difference between efficiencies observed in other works.  A simple analytical model is developed in section~\ref{anal}, allowing the relevant physical processes to the identified.  Section~\ref{apply} discusses the applicability of the results to experiments, and conclusions are drawn in section~\ref{conc}.

\section{Influence of probe duration far beyond the wavebreaking limit}\label{probe}

We make use of the Leap code \cite{raman-farmer-leap}, which is based on a laser-envelope particle-in-cell model.  Since the Raman interaction is predominantly planar due to the short wavelength of the excited plasma wave and the relatively large interaction cross section, we here limit ourselves to one-dimensional (1D) simulations.  In addition to a low computational overhead, this geometry has the further advantage that the plasma response modelled by the Leap code is exact \cite{raman-farmer-envelope}.  Although multidimensional simulations are certainly important for direct comparison to experiments, the fundamental processes are often obfuscated, as different regimes of amplification may be in effect at different radii, discussed in more detail in section~\ref{apply}.  We therefore make use of 1D simulations in order to characterise the coherent-wavebreaking regime of interest here.

\begin{figure*}[t]
 \center{
\includegraphics{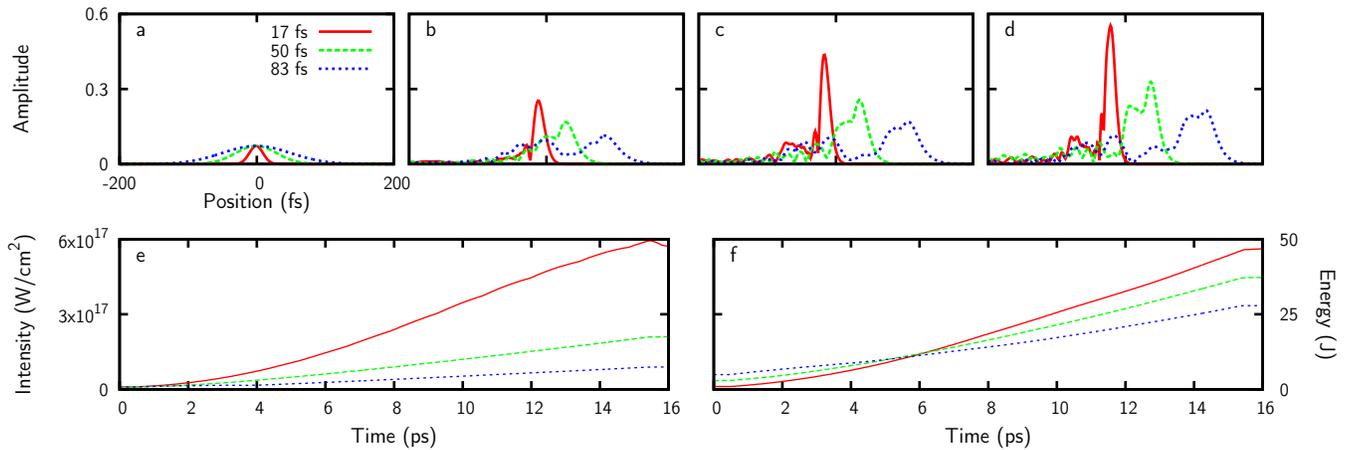}
 }
 \caption{\label{evolve} (Color online) Plots showing evolution of the probe pulse for three different initial durations.  a-d) Snapshots of the amplitude at 0.5, 5.5, 10.5 and 15.5~ps.  e) Peak probe intensity against time.  f) Total probe energy against time.}
\end{figure*}

Figure~\ref{evolve} shows the evolution of the probe pulse for different initial probe durations.  Gaussian pulses are used, with FWHM-intensity durations of 17, 50 and 83~fs\c{, corresponding to sigma of the amplitude of 10, 30 and 50~fs}.  A flat-top, 800~nm pump with duration 30~ps and intensity 1\E{15}\wcm\ is used to amplify a probe with initial intensity 1\E{16}\wcm\ in a plasma of density 4.4\E{18}\ccm\ with an initial temperature of 10~eV.  This corresponds to $\omega_a/\omega_p=20$, $\omega_b/\omega_p=19$,\c{with $\omega_{a,b,p}$ the pump, probe and plasma frequencies} and a pump intensity 30 times above the threshold for wavebreaking to occur \cite{raman-malkin-pumpdepletion}.  
These are the same parameters as used by Toroker~\etal\ \cite{raman-toroker-break} in their investigation of the strong-wavebreaking regime, with the longest pulse used here equivalent to the probe used in that work.  Although the longer pump length used here (30~ps compared to $\sim$1~ps) would result in higher plasma temperatures, we retain the 10~eV temperature to allow direct comparison to those results.  This longer pump allows comparison with the results of Trines~\etal

Figures~\ref{evolve}a-d show snaptshots of the probe profiles at different times during the interaction, given as the reduced vector potential $eE/mc\omega_b$, with $E$ the electric field amplitude, $c$ the vacuum speed of light, and $-e$ and $m$ the electron charge and mass.  It can be seen that in all cases the probe is amplified and compressed.  However, as highlighted in Fig.~\ref{evolve}e, the greatest amplification is achieved for the shortest pulse, which reaches an intensity of 6\E{17}\wcm.  The reduction in peak intensity at the end of the simulation is due to dispersion as the probe continues to propagate through plasma after the end of the pump at 15.5~ps.  Energies were calculated by scaling to three dimensions, taking a pulse radius (e-squared-folding distance) of 595\mum, equal to that used by Trines \etal\ \cite{raman-trines-simulation}, and are shown in Fig.~\ref{evolve}f.  
While the 17~fs probe is initially the least energetic and is compressed to 9.7~fs during amplification, it goes on to exceed the energy of the longer pulses.  The total pump energy for these parameters is 166~J, giving pump-to-probe energy-transfer efficiencies over the full interaction for the three probe durations of, from shortest to longest, 28, 21 and 14\%.

We note from Fig.~\ref{evolve}f that the growth of the probe energy is better than linear, with the efficiency increasing over the interaction as the probe becomes more intense.  The gain of 23~J over a 30~ps pump observed for the 83~fs probe is therefore consistent with the results of Toroker~\etal, which for this pulse width correspond to an energy-transfer of $\sim$0.5~J over $\sim$1~ps pump.  
Trines~\etal\ used similar parameters to the 50~fs probe used here, but with cold plasma and an on-axis intensity for pump and probe double that used here.  However, even with those factors taken into account, the amplification here remains somewhat lower than observed in that work.  This disparity could be a result of the initial pulse shape - using a pulse profile with a finite support, for example truncating a Gaussian at $2\sigma$, increases the amplification, giving comparable results.

The improved amplification observed for shorter probe pulses and the sensitivity to the pulse shape are linked to self-steepening, which causes the peak of the amplified pulse to move forward, away from the initial maximum.  The effective seed amplitude is therefore reduced, as a point on the leading edge of the initial pulse becomes the seed for amplification, rather than than the peak.

\begin{figure*}[t]
 \center{
\includegraphics{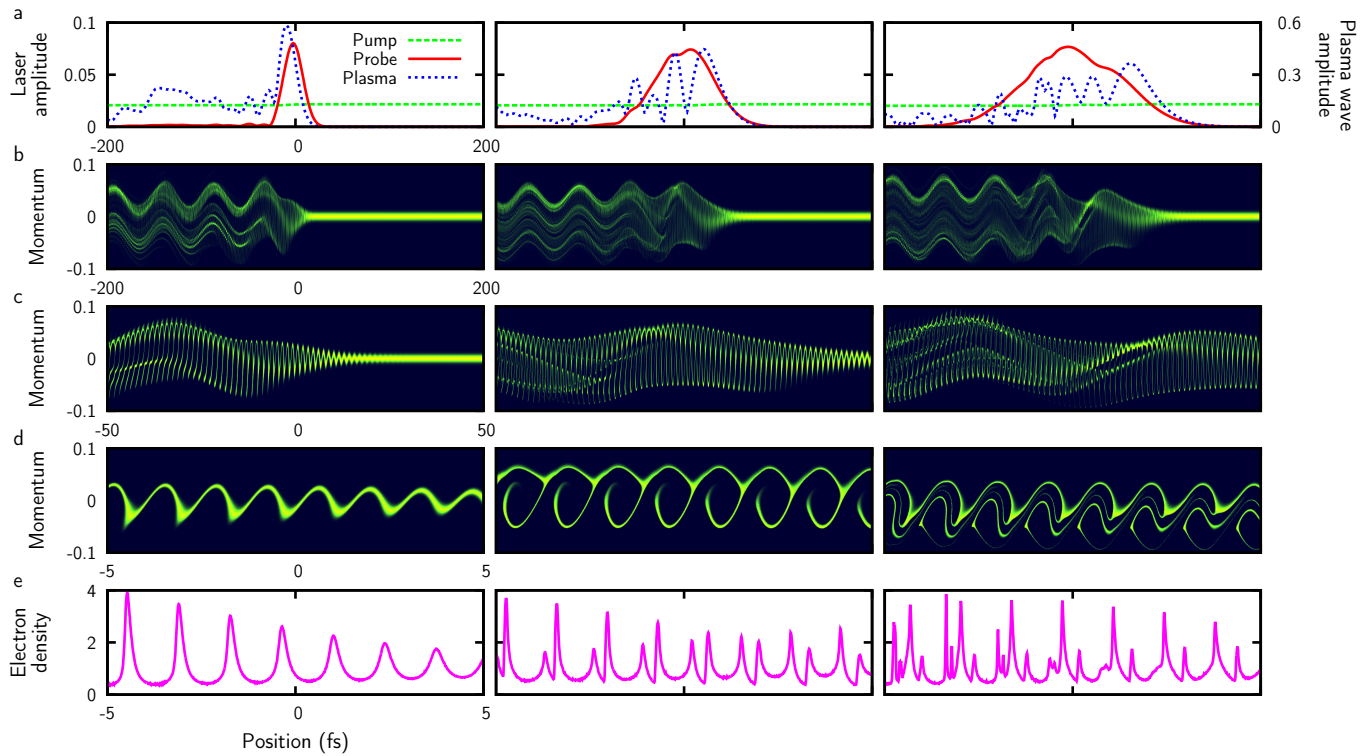}
}
\caption{\label{phase} (Color online) Snapshot of the Raman interaction at time 1.2~ps for initial probe durations of (from left to right) 17, 50 and 83~fs.  a) The pump and probe amplitudes and the effective coupling between the two.  b-d) plots of the electron phase space on different scales, showing b) the long-wavelength oscillation excited by the breaking plasma wave, c) the onset of wavebreaking itself ($v_\mathrm{br}\approx0.025c$), and d) the electron phase space at the probe peak.  e) shows the corresponding electron density at the probe peak.}

\end{figure*}

Self-steepening in the Raman process can arise due to many different effects, such as ponderomotive nonlinearity \cite{raman-shvets-compton}, pump depletion \cite{raman-malkin-pumpdepletion} and the use of a chirped pump \cite{raman-ersfeld-chirp}, while the presence of a pre-pulse can lead to superluminous precursors \cite{raman-tsidulko-precursor}.  
The self-steepening observed here is a result of wave breaking, as can be seen in Fig.~\ref{phase}.  Snapshots of the interaction 1.2 ps into the simulation, before significant amplification occurs, show the pump, probe, and plasma-wave amplitude, the electron phase space and the electron density.  The plasma wave amplitude is here given as the absolute value of the coupling susceptibility \cite{raman-farmer-leap} normalised to the square of the plasma frequency, $|\tilde{\psi}|/\omega_p^2=|\left<(n/\gamma n_0) \e^{\i(\phi_a-\phi_b)}\right>|$, where $\phi_{a,b}$ are the vacuum phases of the pump and probe, $n$ and $n_0$ are the local and equilibrium plasma electron densities, and $\gamma$ the plasma-electron Lorentz factor.  This value represents the coupling between pump and probe.  From linear theory \cite{plasma-dawson-nonlinear}, wavebreaking occurs when the electron velocity exceeds the phase velocity of the wave, $v_\textrm{br}/c\approx\omega_p/2\omega_a=0.025$, corresponding to 
$|\tilde{\psi}|/\omega_p^2=0.5$.

For the longer pulses, wavebreaking occurs on the leading edge of the probe.  As a result, phase mixing of the plasma wave has already set in by the time the probe peak is reached, as seen in Fig.~\ref{phase}d.  The resulting density perturbation, shown in Fig.~\ref{phase}e, loses its periodic structure and no longer efficiently scatters the pump into the probe, lowering amplification.\c{Although the magnitude of the coupling susceptibility oscillates after wavebreaking, similar to a plasma echo as electrons make synchrotron-like oscillations in the broken wave, subsequent peaks have the wrong phase to resonantly scatter the pump into the probe.  Efficient amplification therefore relies on the energy transfer in the first peak, before phase-mixing occurs.  This is the case for the shortest pulse in Fig.~\ref{phase}, in which peak coupling almost coincides with the peak of the probe, leading to maximum amplification at this point.}

Although the electrons periodically rephase as they make synchrotron-like oscillations in the broken wave, the shift in resonance means that these plasma echoes have the wrong phase to coherently scatter the pump into the probe.  Amplification therefore depends on the coupling susceptibility prior to phase mixing.  For the shortest pulse in Fig.~\ref{phase}, the first peak in the coupling susceptibility almost coincides with the peak of the probe, maximising the energy transfer.

If the probe is too short, however, peak amplification may occur behind the probe peak, again lowering energy transfer.  This is in fact the case for the 17~fs probe for time $<1.2$~ps, giving a lower energy-transfer rate than observed for the 83~fs probe.  However, as the probe is amplified, the point at which wavebreaking occurs moves forwards.  This causes the point of peak coupling to advance, leading to improved overlap with the probe, increasing energy transfer.  From 1.2~ps onwards, the energy-transfer rate is highest for the 17~fs probe, which goes on to become the most energetic after $\sim 5.5$~ps.

In addition to the short-wavelength excitation, ${\sim\ifrac{\pi c}{\omega_a}}$, driven by the beat of the pump and probe, it is interesting to note that there is a long-wavelength excitation, ${\sim\ifrac{2\pi c}{\omega_p}}$, as seen in Fig.~\ref{phase}b.  This is not a wake driven by the probe, and is not observed below the wavebreaking threshold, and is instead driven by the breaking of the short-wavelength wave.  From 1D cold-plasma theory, below the wave-breaking threshold, individual electron charge sheets do not cross.  The force acting on a sheet therefore depends only on the smeared-out ionic background, leading to simple harmonic motion\cite{plasma-dawson-nonlinear}.  When the wave breaks, charge sheets cross, and so experience an additional force due to the electron-charge imbalance.  Since the point at which the wave breaks travels with the probe, the associated plasma excitation has a phase velocity ${\sim c}$, resulting in a long-wavelength excitation.

A similar effect has been identified in low-density plasma, in which the excitation arises from ponderomotive nonlinearity, and has been suggested as a method to allow controlled electron acceleration with relatively low-intensity pulses.\cite{wakefield-shvets-periodicstructures}  The effect observed here has the potential advantage that higher density plasma may be used, allowing a larger accelerating field.  Further, there is no constraint on the probe duration, making the effect more widely accessible; for a sufficiently intense pump, the only requirement is that the integrated probe amplitude is sufficiently high for wavebreaking to occur.

\section{Analytical model}\label{anal}
The decrease in the effective probe amplitude caused by self-steepening is comparable to the effect of shadowing \cite{raman-yampolsky-shadow}, in which the leading edge of the probe is preferentially amplified, lowering the effective seeding power.  The effect here, however, is distinct, as wavebreaking changes the interaction, preventing the $\pi$-pulse solution which arises in the pump-depletion regime \cite{raman-malkin-pumpdepletion}.  To illustrate this, we consider the coupled three-wave equations widely used in the study of parametric amplification.  Phenomenological treatments for wavebreaking in the three-wave model have been investigated in other works \cite{raman-balakin-break,raman-farmer-pop}, as have three-wave models incorporating damping and frequency shifts calculated from a nonlinear density distribution function \cite{raman-lindberg-etw}.  Rather than attempting to derive the exact behaviour of the system in regimes far above the wavebreaking threshold, we find that using 
only simple assumptions we can recover the qualitative behaviour observed in Fig.~\ref{evolve}.  We note that, although dispersion plays a role towards the end of the simulation in Fig.~\ref{evolve}, amplification remains the dominant process until the end of the pump-probe interaction.  We therefore limit ourselves to the linearised, dispersion-free equations for pump, probe and plasma wave, to obtain:
\begin{align}
\left(\dt-\dz\right) a &= \frac{\i}{2\omega_a} \tilde{\psi}^\ast b,\nonumber\\
\left(\dt+\dz\right) b &= \frac{\i}{2\omega_b} \tilde{\psi} a,\nonumber\\
\left(\dt + \Omega\right) \tilde\psi &= \frac{\i\omega_p\omega_a^2}{2} a^\ast b.
\end{align}
Here $a$ and $b$ are the envelopes of the pump and probe, respectively, which satisfy $\vec{a} = \real\left(\left(a\e^{\i\phi_a}+b\e^{\i\phi_b}\right)\vec{u}\right)$, with $\vec{a}$ the reduced vector potential, $\phi_{a,b}=\omega_{a,b}(t\pm z/c)$ the carrier phases of the pump and probe, and $\vec{u}$ the polarisation vector, $\vec{u}=(\hat{x}+\i\hat{y})/\surd{2}$ for circularly polarised light.  We retain the coupling susceptibility, $\tilde\psi=\left<(ne^2/\varepsilon_0\gamma m)\e^{\i(\phi_a-\phi_b)}\right>$, which is related to the normalised electric field used in, e.g., \cite{raman-malkin-pumpdepletion}, by $\tilde\psi=2\i\omega_a\omega_pf^\ast$.  The functional $\Omega[x,t,\tilde\psi]$ allows the effects of wavebreaking to be taken into account.  
Assuming an initially cold plasma, valid for the low temperatures considered here, we expect no influence from wavebreaking below the threshold $\psi_\textrm{br}=0.5\omega_p^2$, which yields:
\begin{align}
  \Omega[x,t,\tilde\psi] =
  \begin{cases}
    0,\qquad& \max\limits_{0\leq t^\prime\leq t}\left(|\tilde{\psi}(x,t^\prime)\right|)\leq \tilde\psi_\textrm{br},\\
    \nu+\i\delta,\qquad& \max\limits_{0\leq t^\prime\leq t}\left(|\tilde{\psi}(|x,t^\prime)|\right)>\psi_\textrm{br},
  \end{cases}    
\end{align}
where $\nu$ and $\delta$ are the damping rate and shift in the resonant frequency of the plasma wave due to wavebreaking (in a full treatment, these quantities will certainly vary in time).

Assuming the probe is sufficiently intense for wavebreaking to occur (readily satisfied for high pump intensities), the analytical $\pi$-pulse solution for the pump-depletion regime will remain valid only for the leading edge of the interaction, before the wave breaks.  Substituting the wavebreaking threshold for $\tilde\psi$ into these equations, we obtain the pump amplitude at wavebreaking, $a_\textrm{br}=(a_0^2-(\omega_p/2\omega_a)^3)^{1/2}$.  We note that $a_\textrm{br}$ is independent of the initial probe amplitude.  Moreover, we find that the pump depletion at wavebreaking, $\sim a_0^2-a_\textrm{br}^2=(\omega_p/2\omega_a)^3$, is independent of both pump and probe amplitudes, and corresponds to the threshold pump intensity for wavebreaking to occur.

The energy density of the plasma wave at wavebreaking depends only on its frequency and wavenumber.  Since the pump depletion is fixed, it follows that the intensity increase of the probe up to wavebreaking is also independent of the laser amplitudes.  Either the probe growth rate is high, in which case wavebreaking occurs rapidly, or the growth rate is low, and wavebreaking occurs proportionally later.  The energy gain can only be modified by changing the laser and plasma frequencies, which alters the energy partition and the wavebreaking threshold, or by changing the interaction volume, i.e. a longer pump or wider-diameter beams.

Therefore, in order to improve amplification, we must consider the interaction after wavebreaking.  As the plasma electrons move with finite velocity, $\tilde\psi$ must be continuous in time, and so $\Omega$ must remain finite.  Since the damping and the shift in resonance arise due to phase mixing within the broken wave, we expect the characteristic decay time of the wave to be of the order $1/\omega_p$.  In the finite time for coupling to break down after wavebreaking, we therefore find the total energy transfer will be greater for larger pump and probe amplitudes.

Put simply, the energy transfer prior to wavebreaking is independent of the laser amplitudes, and so efficient amplification relies on maximising the energy transfer after the wave has broken.  This is achieved by maximising the probe amplitude in the period immediately after wavebreaking, before phase mixing destroys the coupling between pump and probe, which requires high-contrast probe pulses of duration ${\sim1/\omega_p}$.

The ``soft-wavebreaking'' model discussed by Balakin \cite{raman-balakin-break}, in which wavebreaking limits but does not reduce the probe growth, can be considered as equivalent to this model in the limit of very low plasma density, where the probe duration is $\ll 1/\omega_p$.  The model discussed here, however, remains valid for the higher plasma densities necessary for significant amplification.

We therefore make a distinction between the wavebreaking regime \cite{raman-toroker-break}, in which amplification of a long probe is significantly reduced by phase mixing after the plasma wave breaks, and the coherent wavebreaking (CWB) regime, observed for the shortest pulse in Fig.~\ref{evolve}, in which the probe is sufficiently short that the probe peak is amplified, even after the wave breaks.  The CWB regime is therefore characterised by much higher efficiencies than the wavebreaking regime, and so the use of a high-quality probe pulse, with short duration and good contrast ratio, is vital for experimental campaigns.

\section{Applicability to experiments}\label{apply}
\begin{figure}
 \center{
\includegraphics{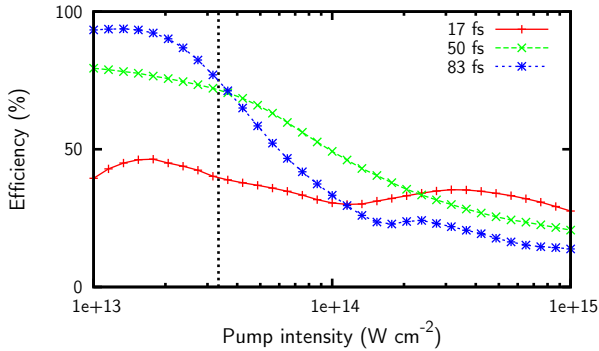}
}
\caption{\label{efficiency} (Color online) Pump-to-probe energy-transfer efficiency for varying pump intensity.  The vertical dashed line shows the wavebreaking threshold.}
\end{figure}

To illustrate this importance, Fig.~\ref{efficiency} shows the pump-to-probe energy-transfer efficiency while varying the pump intensity for different initial probe durations.  Only energy transferred to the probe within 500~fs the initial probe peak is considered.  Other parameters are as for Fig.~\ref{evolve}.  Using a shorter pulse gives rise to higher efficiencies over a wide range of pump intensities beyond the wavebreaking threshold.  Despite this, better efficiencies are achieved by using a pump below the wavebreaking threshold.  In this regime, longer probe pulses exhibit higher efficiencies, as the more energetic probe is better able to deplete the pump.

Although below-wavebreaking amplification yields higher efficiencies, this regime is not necessarily preferable for experiments, as the large interaction volumes required for the same total pump energy may be technically difficult to achieve.  Physically, changing the pump intensity while keeping the probe constant, as is the case for Fig.~\ref{efficiency}, can be understood as the effect of stretching the pump in time -- the same total energy for a 1\E{13}\wcm\ pump would therefore require a 45~cm interaction length.  For limited interaction volumes, then, a higher intensity pump may be preferable in order to maximise energy transfer, despite the decrease in efficiency.  With these considerations in mind, the use of a shorter probe pulse becomes important in order to access the the CWB regime.  This regime has the additional advantage that the duration of the amplified probe can also be significantly shorter than that achieved at lower pump intensities.  

Although the $\sim$95\% efficiency observed at low pump intensity is more than one order of magnitude better than the best experimental results to date, we note that no experiments have been carried out with similar parameters -- Ren~\etal\ \cite{raman-ren-experiment} used an initial probe intensity $\sim$\ten{12}\wcm, compared to the 1\E{16}\wcm\ used here.  We note, however, that the low efficiencies observed in recent campaigns at petawatt-scale facilities (Vulcan, PHELIX) may be due to the lack of a suitable probe -- the shortest available pulses at such facilities are often $\gg100$~fs duration.

Multidimensional simulations show that for beams with Gaussian transverse profiles, the optimal probe duration for a given pump intensity tends to be longer than that shown here, although the general trends are the same.  This is due to the fact that at larger radii, the pump and probe intensities are lower, and so while the centre of the interaction may be far beyond wavebreaking, some parts of the interaction will be in the near-wavebreaking or below-wavebreaking regimes, for which longer probe durations are preferable.  As each regime has different requirements for optimal efficiency, purpose-built Raman amplifiers would likely benefit from the use of flat-top profile beams, as the optimal probe parameters are the same for all radii.

\section{Conclusions}\label{conc}
To conclude, we identify the coherent wavebreaking (CWB) regime of Raman amplification, in which significant amplification occurs after the onset of wavebreaking.  The regime is accessed by using a short, intense probe pulse, which results in wavebreaking occurring close to the peak of the amplified probe, maximising the energy transfer from pump to probe in the time before significant phase mixing occurs.  This regime is of great importance to achieving high amplification, as experimental constraints act to limit the possible interaction volume, and as such limit the possible energy-transfer that can be attained below the wavebreaking threshold.

\begin{acknowledgments}This work was funded by DFG TR18, EU FP7 EuCARD$^2$ and BMBF.
\end{acknowledgments}


%

\end{document}